\documentclass[usegraphicx]{mn2e}

\title[The effective stability parameter for two-component galactic discs]
      {The effective stability parameter for two-component galactic discs:
       Is $Q^{-1}\approx Q_{\mathrm{stars}}^{-1}+Q_{\mathrm{gas}}^{-1}$ ?}

\author[A. B. Romeo and J. Wiegert]
       {Alessandro B. Romeo$^{1}$\thanks{E-mail: romeo@chalmers.se}
        and Joachim Wiegert$^{2}$\\
        $^{1}$Onsala Space Observatory,
              Chalmers University of Technology,
              SE-43992 Onsala, Sweden\\
        $^{2}$Department of Physics,
              University of Gothenburg,
              SE-41296 Gothenburg, Sweden}

\begin{document}

\date{Accepted 2011 May 21.
      Received 2011 April 26; in original form 2011 January 24}

\pagerange{\pageref{firstpage}--\pageref{lastpage}}

\pubyear{2011}

\maketitle

\label{firstpage}

\begin{abstract}
The Wang-Silk approximation, $Q^{-1}\approx
Q_{\mathrm{stars}}^{-1}+Q_{\mathrm{gas}}^{-1}$\,, is frequently used for
estimating the effective $Q$ parameter in two-component discs of stars and
gas.  Here we analyse this approximation in detail, and show how its accuracy
depends on the radial velocity dispersions and Toomre parameters of the two
components.  We then propose a much more accurate but still simple
approximation for the effective $Q$ parameter, which further takes into
account the stabilizing effect of disc thickness.  Our effective $Q$
parameter is a natural generalization of Toomre's $Q$, and as such can be
used in a wide variety of contexts, e.g.\ for predicting star formation
thresholds in galaxies or for measuring the stability level of galactic discs
at low and high redshifts.
\end{abstract}


\section{INTRODUCTION}

It is well known that both stars and cold interstellar gas play an important
role in the gravitational instability of galactic discs (e.g., Lin \& Shu
1966; Jog and Solomon 1984a,\,b; Bertin \& Romeo 1988, and references
therein).  The local stability criterion for two-component discs of stars and
gas can be expressed in the same form as Toomre's (1964) stability criterion,
$Q\geq1$, provided that $Q$ is redefined appropriately.  Bertin \& Romeo
(1988), Elmegreen (1995), Jog (1996), Rafikov (2001) and Shen \& Lou (2003)
calculated the effective $Q$ parameter as a function of the radial velocity
dispersions and surface mass densities of the two components.  Shu (1968),
Romeo (1990, 1992, 1994) and Wiegert (2010) evaluated the stabilizing effect
of disc thickness, which is usually neglected but significant.

As galactic discs contain both stars and gas, the effective $Q$ parameter is
clearly more accurate and useful than Toomre's $Q$.  Bertin et
al.\ (1989a,\,b) and Lowe et al.\ (1994) showed that the radial profile of
this parameter has a large impact on the dynamics and evolution of spiral
structure in galaxies.  The results of those comprehensive analyses are
discussed further in the books by Bertin \& Lin (1996) and Bertin (2000).
The effective $Q$ parameter is also a useful diagnostic for exploring the
link between disc instability and star formation in galaxies (e.g., Hunter et
al.\ 1998; Li et al.\ 2005, 2006; Yang et al.\ 2007; Leroy et al.\ 2008).
More applications and references are given below.

Wang \& Silk (1994) proposed a remarkably simple recipe for computing the
effective $Q$ parameter in the case of infinitesimally thin discs:
$Q^{-1}\approx Q_{\star}^{-1}+Q_{\mathrm{g}}^{-1}$, where $Q_{\star}$ and
$Q_{\mathrm{g}}$ are the stellar and gaseous Toomre parameters.  Bertin
(private communication) points out that such an approximation was already
used by him, before the 1990s, for illustrating the efficiency of a small
amount of cold gas to destabilize a disc (see also Bertin 1996; Bertin \& Lin
1996).  Jog (1996) pointed out that the Wang-Silk approximation is invalid
since it results from an incorrect analysis.  In spite of that, the Wang-Silk
approximation has been used in several important contexts: star formation
(e.g., Martin \& Kennicutt 2001; Boissier et al.\ 2003; Corbelli 2003; Wong
2009), galaxy formation and evolution (e.g., Immeli et al.\ 2004; Naab \&
Ostriker 2006; Kampakoglou \& Silk 2007; Stringer \& Benson 2007; Wetzstein
et al.\ 2007; Foyle et al.\ 2008; Benson 2010), gravitational instability of
clumpy discs at low and high redshifts (e.g., Bournaud \& Elmegreen 2009;
Burkert et al.\ 2010; Puech 2010), and others (e.g., Hitschfeld et al.\ 2009;
Wong et al.\ 2009).

In spite of such a burst of applications, there has been no attempt to assess
how good the Wang-Silk approximation is.  In this paper, we evaluate its
accuracy by performing a rigorous comparative analysis (see Sect.\ 2.1).
Besides, we introduce a new approximation for the effective $Q$ parameter:
simple, accurate and applicable to realistically thick discs (see Sects 2.2
and 2.3).  We also show how to use our effective $Q$ parameter for measuring
the stability level of galactic discs, and why such a diagnostic is more
predictive than the classical Toomre parameter (see Sect.\ 2.4).  The
conclusions of our paper are drawn in Sect.\ 3.

\section{APPROXIMATING THE EFFECTIVE $Q$}

\subsection{Disc instability and \hspace{4cm}
            the Wang-Silk approximation}

\begin{figure}
\includegraphics[scale=1.]{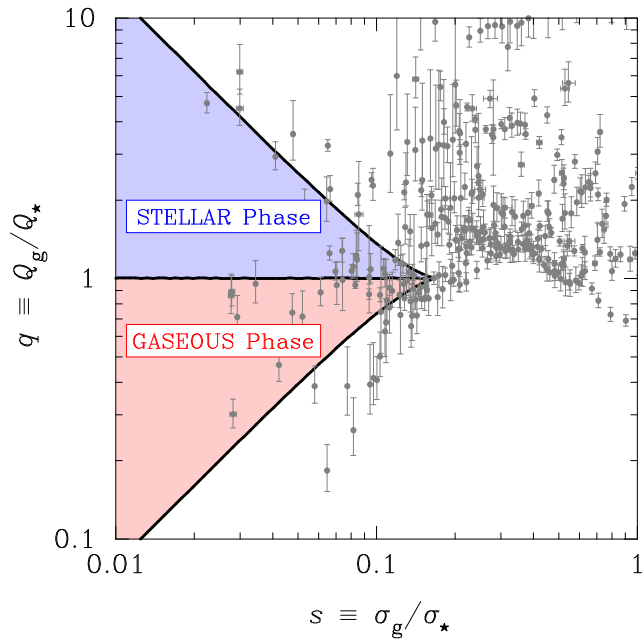}
\caption{The parameter plane populated by nearby star-forming spirals.  The
  galaxy data are from Leroy et al.\ (2008), $Q_{\star}$ and $Q_{\mathrm{g}}$
  are the stellar and gaseous Toomre parameters, $\sigma_{\star}$ and
  $\sigma_{\mathrm{g}}$ are the radial velocity dispersions of the two
  components.  The shaded part of the $(s,q)$ plane is the two-phase region
  discussed in the text.  The dispersion relation $\omega^{2}(k)$ has two
  minima inside this region, and one minimum outside it.  The transition
  between the gaseous and stellar stability phases occurs for $q=1$.  This
  line intersects the boundaries of the two-phase region at
  $(s,q)\simeq(0.17,1)$, where the stability threshold is
  $\overline{Q}_{\mathrm{BR}}\simeq1.4$.}
\end{figure}

\begin{figure*}
\includegraphics[scale=.96]{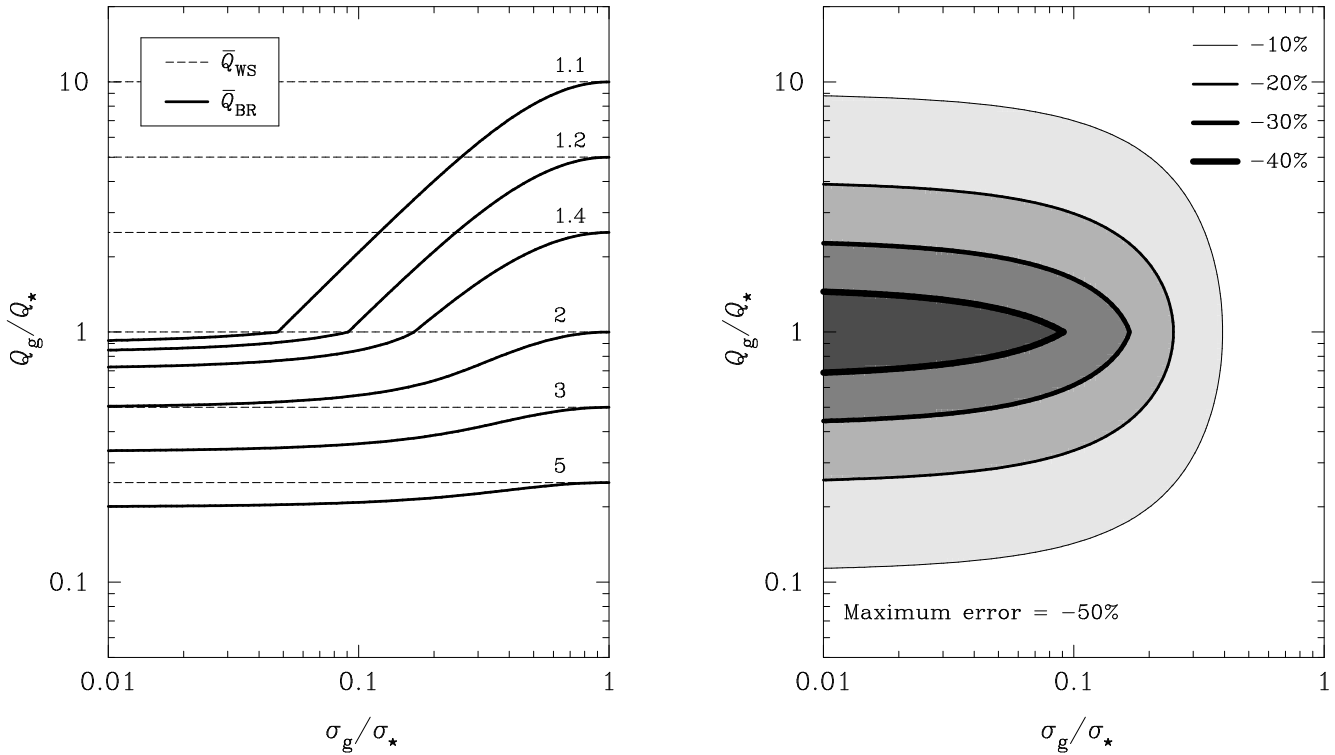}
\caption{Accuracy of the Wang-Silk approximation.  The curves shown are the
  contour lines of $\overline{Q}_{\mathrm{WS}}$ and
  $\overline{Q}_{\mathrm{BR}}$ (left), and the contour lines of the relative
  error $(Q_{\mathrm{WS}}-Q_{\mathrm{BR}})/Q_{\mathrm{BR}}$ (right).  $Q$ and
  $\overline{Q}$ denote the effective $Q$ parameter and the stability
  threshold, the subscripts WS and BR refer to Wang \& Silk (1994) and Bertin
  \& Romeo (1988).  In addition, $Q_{\star}$ and $Q_{\mathrm{g}}$ are the
  stellar and gaseous Toomre parameters, $\sigma_{\star}$ and
  $\sigma_{\mathrm{g}}$ are the radial velocity dispersions of the two
  components.  Note how far the Wang-Silk approximation is from the correct
  stability threshold, except for $\sigma_{\mathrm{g}}\sim\sigma_{\star}$
  (see left panel).}
\end{figure*}

Wang \& Silk (1994) investigated the link between star formation and disc
instability in galaxies.  They reconsidered the two-fluid dispersion relation
of Jog \& Solomon (1984a), which is valid for infinitesimally thin discs of
stars and gas, and found that the effective $Q$ parameter can be approximated
as follows:
\begin{equation}
\frac{1}{Q_{\mathrm{WS}}}=
\frac{1}{Q_{\star}}+\frac{1}{Q_{\mathrm{g}}}\,,
\end{equation}
where $Q_{\star}=\kappa\sigma_{\star}/\pi G\Sigma_{\star}$ and
$Q_{\mathrm{g}}=\kappa\sigma_{\mathrm{g}}/\pi G\Sigma_{\mathrm{g}}$ are the
stellar and gaseous Toomre parameters.  This approximation is appealing
because it is as simple as the formula for the total resistance in a parallel
circuit.  To evaluate the accuracy of Eq.\ (1), we rewrite it as
$Q_{\mathrm{WS}}=Q_{\star}/\,\overline{Q}_{\mathrm{WS}}$, where
\begin{equation}
\overline{Q}_{\mathrm{WS}}=
1+\frac{Q_{\star}}{Q_{\mathrm{g}}}\,.
\end{equation}
The local stability criterion, $Q_{\mathrm{WS}}\geq1$, translates into
$Q_{\star}\geq\overline{Q}_{\mathrm{WS}}$.  This is of the same form as the
local stability criterion found by Bertin \& Romeo (1988); see also Romeo
(1985).  Bertin \& Romeo (1988) determined the stability threshold
$\overline{Q}_{\mathrm{BR}}$ numerically, starting from the same dispersion
relation as Wang \& Silk (1994) but without introducing further
approximations.  In contrast to $\overline{Q}_{\mathrm{WS}}$,
$\overline{Q}_{\mathrm{BR}}$ depends on two parameters:
$\sigma_{\mathrm{g}}/\sigma_{\star}$ and
$\Sigma_{\mathrm{g}}/\Sigma_{\star}$.  Since
$\Sigma_{\mathrm{g}}/\Sigma_{\star}=
\sigma_{\mathrm{g}}Q_{\star}/\sigma_{\star}Q_{\mathrm{g}}$, we can easily
express $\overline{Q}_{\mathrm{BR}}$ in terms of
\begin{equation}
s\equiv\frac{\sigma_{\mathrm{g}}}{\sigma_{\star}}\,,\;\;\;
q\equiv\frac{Q_{\mathrm{g}}}{Q_{\star}}\,.
\end{equation}
We can then compare $\overline{Q}_{\mathrm{WS}}(q)$ with
$\overline{Q}_{\mathrm{BR}}(s,q)$ and evaluate the accuracy of the Wang-Silk
approximation as a function of $s$ and $q$.

Let us first see how spiral galaxies populate the $(s,q)$ plane.  We use the
12 nearby star-forming spirals analysed by Leroy et al.\ (2008), namely NGC
628, 2841, 3184, 3198, 3351, 3521, 3627, 4736, 5055, 5194, 6946 and 7331.
These are galaxies with sensitive and spatially resolved measurements of
kinematics, gas surface density and stellar surface density across the entire
optical disc.  For each galaxy of this sample, we compute the radial profiles
$s=s(R)$ and $q=q(R)$, and hence the track left by the galaxy in the $(s,q)$
plane.  The result for the whole sample is shown in Fig.\ 1.  The data span a
range of two orders of magnitude in $s$ and $q$, so we show them using a
Log-Log plot.  The typical value of $s$ can be robustly estimated by
computing the median of the data points along $s$, which is
$s_{\mathrm{med}}\simeq0.27$.  This value is comparable to that found in the
solar neighbourhood ($s\approx0.2$; see Binney \& Tremaine 2008, p.\ 497),
but is much smaller than that expected in high-redshift star-forming galaxies
($s\sim1$; e.g., Burkert et al.\ 2010; Krumholz \& Burkert 2010).  The median
value of $q$, $q_{\mathrm{med}}\simeq1.5$, is close to unity.  This suggests
that, on average, stars and gas contribute equally to the gravitational
instability of the disc.  Similar values of $q$ are found in the solar
neighbourhood ($q\approx0.6$; see Binney \& Tremaine 2008, p.\ 497), and are
also expected at high $z$ ($q\sim1$; e.g., Burkert et al.\ 2010; Krumholz \&
Burkert 2010; Tacconi et al.\ 2010).%
\footnote{Hereafter we will use $s_{\mathrm{med}}$ and $q_{\mathrm{med}}$,
  i.e.\ the median values of $s$ and $q$ computed from the galaxy data of
  Leroy et al.\ (2008), for estimating the typical accuracy of the Wang-Silk
  approximation and of our approximation.  This is meant to be a complement
  to the detailed error maps shown and discussed throughout the paper.  We do
  not `hint that the stability properties can be characterized by a median
  value of an effective $Q$-parameter'.}
Last but not least, note that 20\% of the data fall within the shaded part of
the $(s,q)$ plane.  This is the `two-phase region' of Bertin \& Romeo (1988),
here shown using our parametrization and logarithmic scaling.  In this
region, the contributions of stars and gas to the gravitational instability
of the disc peak at two different wavelengths.  If $q<1$, then the gaseous
peak is higher than the stellar one and gas will dominate the onset of
gravitational instability.  Vice versa, if $q>1$, then stars will dominate.
These are the gaseous and stellar stability `phases' shown in Fig.\ 1.  In
the rest of the parameter plane, the dynamical responses of the two
components are strongly coupled and peak at a single wavelength.  More
information is given in sect.\ 3.2.2 of Romeo (1994).

Fig.\ 2 shows the contour maps of $\overline{Q}_{\mathrm{BR}}$ and
$\overline{Q}_{\mathrm{WS}}$ (left panel), and the error map of
$Q_{\mathrm{WS}}$ (right panel).  Remember that $\overline{Q}$ denotes the
stability threshold, i.e.\ the value of $Q_{\star}$ above which the
two-component disc is locally stable, while $Q=Q_{\star}/\,\overline{Q}$
denotes the effective $Q$ parameter.  Both stability thresholds are above
unity and converge to $1+q^{-1}$ as $s\rightarrow1$.  The first property
means that a disc of stars and gas can be gravitationally unstable even when
both components are separately stable, as is well known (e.g., Lin \& Shu
1966).  The second property simply means that stars and gas act as a single
component when they have the same radial velocity dispersion, so that
$Q=\kappa\sigma/\pi G(\Sigma_{\star}+\Sigma_{\mathrm{g}})$.  Apart from
satisfying those properties, the two stability thresholds are clearly
different.  Look in particular at the contour levels 1.1--1.4 of
$\overline{Q}_{\mathrm{BR}}$.  Their slope changes abruptly across the line
$q=1$, revealing the existence of two stability phases.  As discussed in the
previous paragraph, this is an important characteristic of two-component
discs, which $\overline{Q}_{\mathrm{WS}}$ fails to reproduce.  From a
quantitative point of view, the error that affects $Q_{\mathrm{WS}}$ is
significant but below 50\% (see now the right panel of Fig.\ 2).  Using the
median values of $s$ and $q$ computed from the galaxy data of Leroy et
al.\ (2008), $s_{\mathrm{med}}\simeq0.27$ and $q_{\mathrm{med}}\simeq1.5$,
one finds that the typical error is about 20\%.  Remember, however, that a
significant fraction of the data populate the two-phase region, where the
error can be more than 40\%.  Note also that the error is negative, which
means that the Wang-Silk approximation underestimates the effective $Q$
parameter systematically.

\subsection{Our approximation}

\begin{figure*}
\includegraphics[scale=.96]{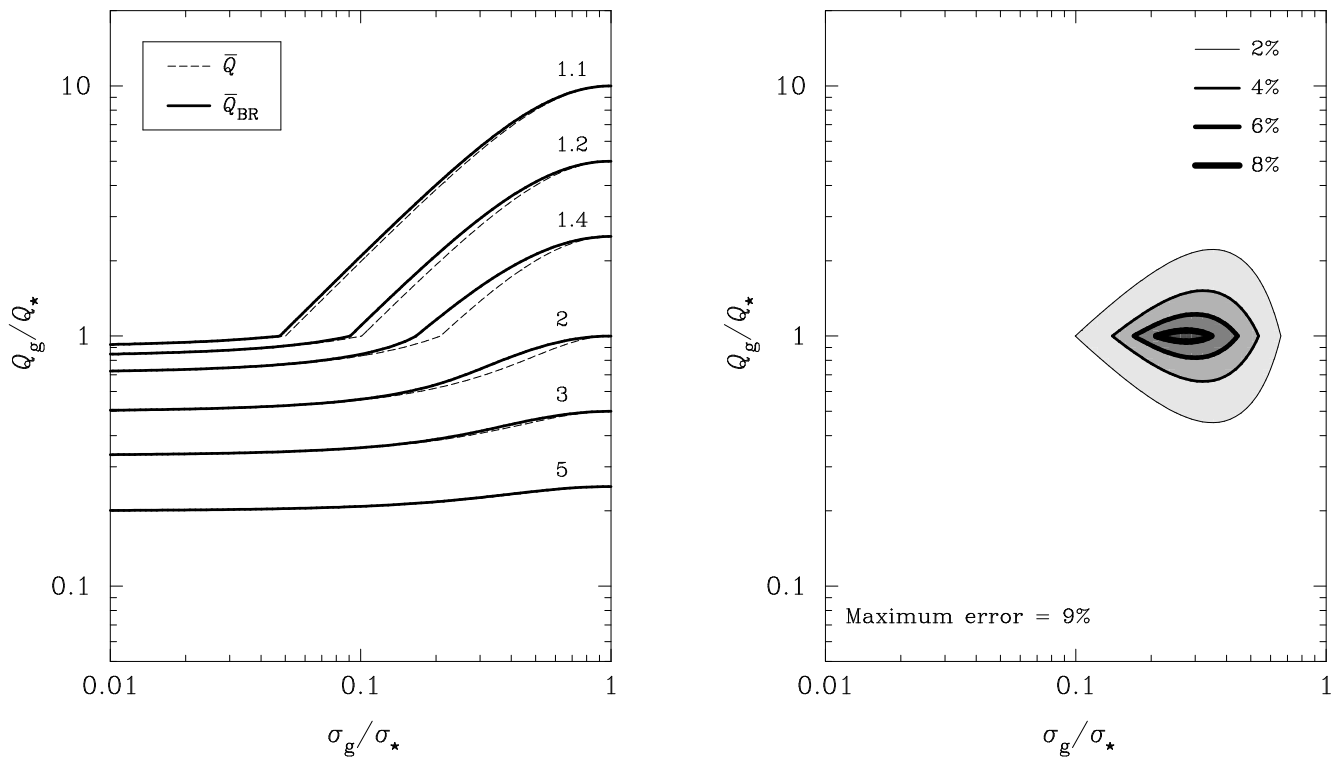}
\caption{Accuracy of our approximation.  The curves shown are the same as in
  Fig.\ 2, but for our effective $Q$ parameter and stability threshold.
  These quantities are denoted by $Q$ and $\overline{Q}$, without subscripts.
  Note how close our approximation is to the correct stability threshold,
  especially for $\overline{Q}\la1.2$ and $\overline{Q}\ga2$ (see left
  panel).}
\end{figure*}

Let us now illustrate how to find a better approximation for $Q$.  The first
ingredient is to determine the asymptotic behaviour of $\overline{Q}$ as
$s\rightarrow0$ and $s\rightarrow1$.  These are in fact the natural bounds of
$s$.  A rigorous analysis was performed by Romeo (1985).  His results can be
summarized as follows:
\begin{enumerate}
\item For $s\ll1$ and $q\leq1$, i.e.\ in gas-dominated stability regimes,
  $\overline{Q}\approx q^{-1}+2s$.
\item For $s\ll1$ and $q\geq1$, i.e.\ in star-dominated stability regimes,
  $\overline{Q}\approx1+2s\,q^{-1}$.
\item For $s\approx1$, i.e.\ in the limiting case of a one-component disc,
  $\overline{Q}\approx1+q^{-1}$.
\end{enumerate}
Note that $\overline{Q}$ behaves asymptotically as a weighted sum of two
terms: 1 and $q^{-1}$.  Note also that the weight factors change
symmetrically as we move from case (i) to case (iii):
$(2s,1)\rightarrow(1,2s)\rightarrow(1,1)$.  Such symmetry suggests that we
should search for an approximation of the form
\begin{equation}
\overline{Q}=
\left\{\begin{array}{ll}
       W(s)+q^{-1}    & \mbox{if\ }q\leq1\,, \\
       1+W(s)\,q^{-1} & \mbox{else}\,;
       \end{array}
\right.
\end{equation}
where $W(s)\approx2s$ as $s\rightarrow0$, and $W(s)\approx1$ as
$s\rightarrow1$.  A further constraint on $W(s)$ follows from the fact that
the original system of fluid and Poisson equations remains unaltered if we
interchange the stellar and gaseous components.  Eq.\ (4) must then be
invariant under the transformation $s\mapsto s^{-1}$, $q\mapsto q^{-1}$ and
hence $\overline{Q}\mapsto q\overline{Q}$, where $q\overline{Q}$ is the value
of $Q_{\mathrm{g}}$ above which the two-component disc is locally stable.
Invariance requires that $W(s^{-1})=W(s)$.  A simple function that satisfies
this requirement and matches the asymptotic behaviour above is
\begin{equation}
W(s)=
\frac{2s}{1+s^{2}}\,.
\end{equation}
Since $Q=Q_{\star}/\,\overline{Q}$, Eqs (4) and (5) lead us to the following
approximation for the effective $Q$ parameter:
\begin{equation}
\frac{1}{Q}=
\left\{\begin{array}{ll}
       {\displaystyle\frac{W}{Q_{\star}}+\frac{1}{Q_{\mathrm{g}}}}
                       & \mbox{if\ }Q_{\star}\geq Q_{\mathrm{g}}\,, \\
                       &                                            \\
       {\displaystyle\frac{1}{Q_{\star}}+\frac{W}{Q_{\mathrm{g}}}}
                       & \mbox{if\ }Q_{\mathrm{g}}\geq Q_{\star}\,;
       \end{array}
\right.
\end{equation}
\begin{equation}
W=
\frac{2\sigma_{\star}\sigma_{\mathrm{g}}}
     {\sigma_{\star}^{2}+\sigma_{\mathrm{g}}^{2}}\,.
\end{equation}
Our approximation is almost as simple as the Wang-Silk approximation
[Eq.\ (1)], but differs from that in one important respect: it gives less
weight to the component with larger $Q$.  The weight factor $W$ depends
symmetrically on the radial velocity dispersions of the two components, and
is generally small.

To evaluate the accuracy of our approximation, we compare $\overline{Q}(s,q)$
with $\overline{Q}_{\mathrm{BR}}(s,q)$ and compute the relative error
$(Q-Q_{\mathrm{BR}})/Q_{\mathrm{BR}}$ as a function of $s$ and $q$, as we did
for the Wang-Silk approximation.  Fig.\ 3 shows that $\overline{Q}$ works
well in the whole parameter space (see left panel).  Note in particular how
successfully our approximation reproduces the gaseous and stellar stability
phases for $\overline{Q}\la1.2$.  Fig.\ 3 also shows that $Q$ overestimates
the effective stability parameter, but the error is well below 10\% even
inside the two-phase region (see right panel).  The error can be reduced
further by fine-tuning the weight factor, but the approximation will no
longer be consistent with the asymptotic behaviour of the stability
threshold.

\subsection{How to apply our approximation to realistically thick discs}

\begin{figure*}
\includegraphics[scale=.96]{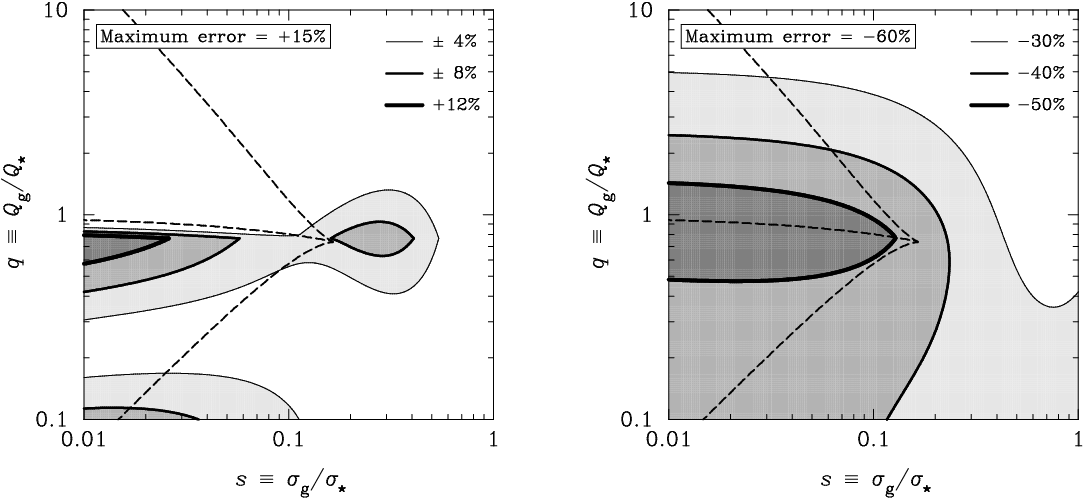}
\caption{Accuracy of our approximation (left) vs.\ accuracy of the Wang-Silk
  approximation (right) for realistically thick discs.  The curves shown are
  the contour lines of the relative errors
  $(\mathcal{Q}-\mathcal{Q}_{\mathrm{WR}})/\mathcal{Q}_{\mathrm{WR}}$ (see
  left panel) and
  $(Q_{\mathrm{WS}}-\mathcal{Q}_{\mathrm{WR}})/\mathcal{Q}_{\mathrm{WR}}$
  (see right panel) for $(\sigma_{z}/\sigma_{R})_{\star}=0.5$ and
  $(\sigma_{z}/\sigma_{R})_{\mathrm{g}}=1$.  Here $\mathcal{Q}$ is our
  effective $Q$ parameter, $\mathcal{Q}_{\mathrm{WR}}$ is the effective $Q$
  parameter of Romeo (1992) and Wiegert (2010), $Q_{\mathrm{WS}}$ is the
  effective $Q$ parameter of Wang \& Silk (1994), and $\sigma_{z}/\sigma_{R}$
  is the ratio of vertical to radial velocity dispersion.  The rest of the
  notation is the same as in Figs 1--3.  Also shown is the corresponding
  two-phase region (dashed lines).  The boundaries of this region and the
  transition line intersect at $(s,q)\simeq(0.16,0.74)$, where the stability
  threshold is $\overline{\mathcal{Q}}_{\mathrm{WR}}\simeq1.3$.}
\end{figure*}

As pointed out in Sect.\ 1, the stabilizing effect of disc thickness is
usually neglected but significant.  In this section, we show that our
approximation can easily be modified so as to take this effect into account.

Romeo (1992) investigated the gravitational instability of galactic discs
taking rigorously into account two factors: (i) their vertical structure at
equilibrium; (ii) the coupling between scale height, $h$, and vertical
velocity dispersion, $\sigma_{z}$, in the stellar and gaseous layers.  He
calculated the effective $Q$ parameter both as a function of $h_{\star}$ and
$h_{\mathrm{g}}$, and as a function of $\sigma_{z\star}$ and
$\sigma_{z\mathrm{g}}$.  He also discussed the advantages of using
$\sigma_{z\star}$ and $\sigma_{z\mathrm{g}}$ as input quantities.  This
effective $Q$ parameter has been studied further by Wiegert (2010).
Hereafter we will denote it with $\mathcal{Q}_{\mathrm{WR}}$.

Let us now illustrate how to find a simple and accurate approximation to
$\mathcal{Q}_{\mathrm{WR}}$.  In the infinitesimally thin case [see
  Eq.\ (6)], the local stability level of the disc is dominated by the
component with smaller $Q$.  The contribution of the other component is
weakened by the $W$ factor, which is generally small.  This suggests that the
effect of thickness can be estimated reasonably well by considering each
component separately.  Romeo (1994) analysed this case in detail.  The effect
of thickness is to increase the stability parameter of each component by a
factor $T$, which depends on the ratio of vertical to radial velocity
dispersion:
\begin{equation}
T\approx
0.8+0.7\left(\frac{\sigma_{z}}{\sigma_{R}}\right)\,.
\end{equation}
Eq.\ (8) can be inferred from fig.\ 3 (top) of Romeo (1994) and applies for
$0.5\la\sigma_{z}/\sigma_{R}\la1$, which is the usual range of velocity
anisotropy.  To approximate $\mathcal{Q}_{\mathrm{WR}}$, use then Eq.\ (6)
with $Q_{\star}$ and $Q_{\mathrm{g}}$ replaced by $T_{\star}Q_{\star}$ and
$T_{\mathrm{g}}Q_{\mathrm{g}}$:
\begin{equation}
\frac{1}{\mathcal{Q}}=
\left\{\begin{array}{ll}
       {\displaystyle\frac{W}{T_{\star}Q_{\star}}+
                     \frac{1}{T_{\mathrm{g}}Q_{\mathrm{g}}}}
                       & \mbox{if\ \ }T_{\star}Q_{\star}\geq
                                      T_{\mathrm{g}}Q_{\mathrm{g}}\,, \\
                       &                                              \\
       {\displaystyle\frac{1}{T_{\star}Q_{\star}}+
                     \frac{W}{T_{\mathrm{g}}Q_{\mathrm{g}}}}
                       & \mbox{if\ \ }T_{\mathrm{g}}Q_{\mathrm{g}}\geq
                                      T_{\star}Q_{\star}\,;
       \end{array}
\right.
\end{equation}
where $\mathcal{Q}$ is our effective $Q$ parameter for realistically thick
discs, $W$ is given by Eq.\ (7), $T_{\star}$ and $T_{\mathrm{g}}$ are given
by Eq.\ (8).  Eq.\ (9) tells us that the local stability level of the disc is
now dominated by the component with smaller $TQ$.  The contribution of the
other component is still suppressed by the $W$ factor.

The left panel of Fig.\ 4 shows the error map of $\mathcal{Q}$ for a galactic
disc with $(\sigma_{z}/\sigma_{R})_{\star}=0.5$ and
$(\sigma_{z}/\sigma_{R})_{\mathrm{g}}=1$, and the corresponding two-phase
region (Wiegert 2010).  Note that the error is below 15\% even inside this
region, which confirms the high accuracy of our approximation in this more
realistic context.  What about the accuracy of the Wang-Silk approximation?
The right panel of Fig.\ 4 shows that the relative error
$(Q_{\mathrm{WS}}-\mathcal{Q}_{\mathrm{WR}})/\mathcal{Q}_{\mathrm{WR}}$ is
much larger than ours, and can be well above 50\% inside the two-phase
region.

\subsection{Application to nearby star-forming spirals}

\begin{figure*}
\includegraphics[scale=.96]{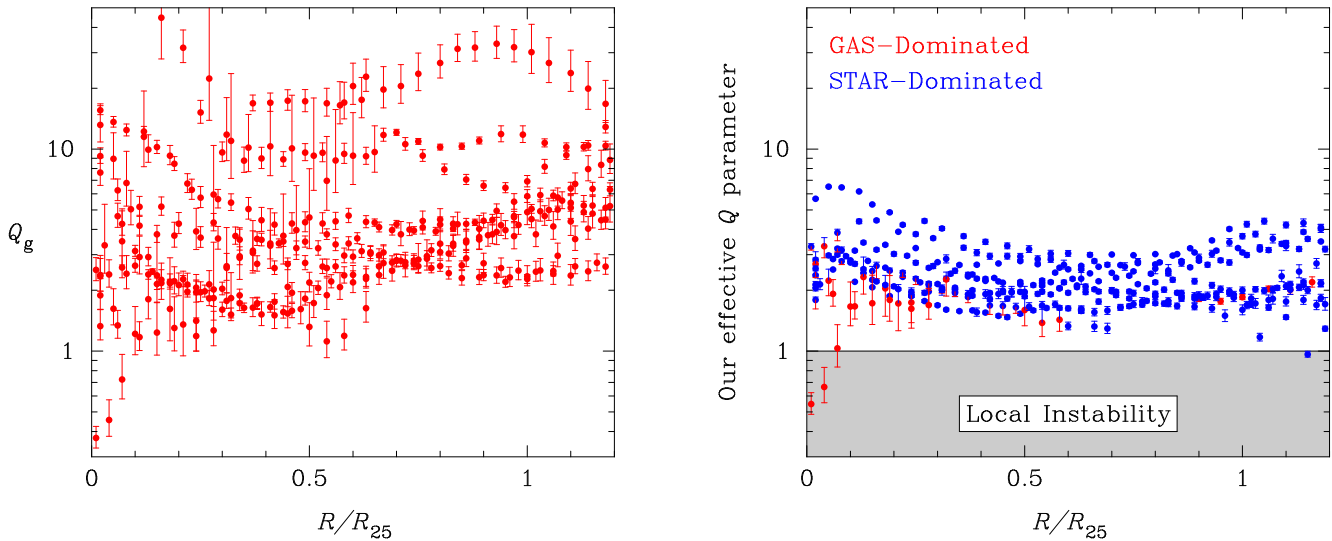}
\caption{The stability level of nearby star-forming spirals, as measured by
  two diagnostics: the gaseous Toomre parameter, $Q_{\mathrm{g}}$, and our
  effective $Q$ parameter, $\mathcal{Q}$ [see Eq.\ (9)].  The galaxy data are
  from Leroy et al.\ (2008), $R$ is the galactocentric distance, and $R_{25}$
  is the optical radius.  In the right panel, the data are colour-coded so as
  to show whether the stability level is gas- or star-dominated, as predicted
  by Eq.\ (9).  The two data points that lie well below the critical
  stability level tell us that the nuclear region of NGC 6946 is subject to
  strong gas-dominated instabilities.  This is consistent with the fact that
  NGC 6946 hosts a nuclear starburst (e.g., Engelbracht et al.\ 1996).}
\end{figure*}

In this section, we show how to use our effective $Q$ parameter for measuring
the stability level of galactic discs, and why such a diagnostic is more
predictive than the classical Toomre parameter.

We consider the same sample of spiral galaxies as in Sect.\ 2.1, and refer to
Leroy et al.\ (2008) for a detailed description of the data and their
translation into physical quantities.  For each galaxy, we compute the radial
profile of our effective $Q$ parameter, $\mathcal{Q}$, using Eq.\ (9).  We
adopt $(\sigma_{z}/\sigma_{R})_{\star}=0.6$, as was assumed by Leroy et
al.\ (2008), and $(\sigma_{z}/\sigma_{R})_{\mathrm{g}}=1$, as is natural for
a collisional component.  We also compute the radial profile of the gaseous
Toomre parameter, $Q_{\mathrm{g}}$, which is the traditional diagnostic used
for predicting star formation thresholds in galaxies (e.g., Quirk 1972;
Kennicutt 1989; Martin \& Kennicutt 2001; Schaye 2008; Elmegreen 2011).

Fig.\ 5 shows $Q_{\mathrm{g}}(R)$ and $\mathcal{Q}(R)$ for the whole galaxy
sample.  Note that $Q_{\mathrm{g}}$ spans a much wider range of values than
$\mathcal{Q}$ at any given $R$.  This is true even at distances as large as
the optical radius, $R_{25}$, where $Q_{\mathrm{g}}$ is supposed to be a
reliable diagnostic.  A similar fact was noted by Leroy et al.\ (2008), using
an effective $Q$ parameter that neglects the stabilizing effect of disc
thickness (Jog 1996; Rafikov 2001).  Why are $Q_{\mathrm{g}}$ and
$\mathcal{Q}$ so weakly correlated across the entire optical disc?  Eq.\ (9)
helps us to clarify this point.  It tells us that the value of $\mathcal{Q}$
is dominated by the gaseous component if
$T_{\mathrm{g}}Q_{\mathrm{g}}<T_{\star}Q_{\star}$, and by the stellar
component if $T_{\star}Q_{\star}<T_{\mathrm{g}}Q_{\mathrm{g}}$.  In the right
panel of Fig.\ 5, we have colour-coded the data so as to show whether
$T_{\mathrm{g}}Q_{\mathrm{g}}<T_{\star}Q_{\star}$ or vice versa.  It turns
out that in 92\% of the cases the value of $\mathcal{Q}$ is dominated by the
stellar component.  Gas dominates the stability level only in 8\% of the
cases.  This is why $Q_{\mathrm{g}}$ and $\mathcal{Q}$ are so weakly
correlated.  This result illustrates (i) how important it is to consider both
gas and stars when measuring the stability level of galactic discs, and (ii)
the strong advantage of using our effective $Q$ parameter as a stability
diagnostic.

\section{CONCLUSIONS}

\begin{itemize}
\item The approximation of Wang \& Silk (1994) [Eq.\ (1)] underestimates the
  effective $Q$ parameter.  The error is typically 20\%, but can be as large
  as 40\% or more if $\sigma_{\mathrm{g}}\la0.2\sigma_{\star}$ and
  $Q_{\mathrm{g}}\sim Q_{\star}$.  In this case, the gaseous and stellar
  components should contribute separately to the gravitational instability of
  the disc (Bertin \& Romeo 1988).  But such dynamical decoupling is
  difficult to approximate because it involves two stability regimes, one
  dominated by the gas and the other dominated by the stars, and because
  there is a sharp transition between the two `phases'.  So it is not strange
  that the Wang-Silk approximation becomes less accurate when
  $\sigma_{\mathrm{g}}\la0.2\sigma_{\star}$ and $Q_{\mathrm{g}}\sim
  Q_{\star}$.
\item Our approximation [Eq.\ (6)] overestimates the effective $Q$ parameter,
  but the error is less than 9\% and typically as small as 4\%.  The accuracy
  and simplicity of our approximation result from a rigorous analysis, which
  takes into account the stability characteristics of the disc as well as the
  symmetries of the problem.
\item We provide a simple recipe for applying our approximation to
  realistically thick discs [see Eq.\ (9)].  The ratio of vertical to radial
  velocity dispersion is usually 0.5 for the stars and 1 for the gas.  In
  this case, our approximation is in error by less than 15\%, whereas the
  Wang-Silk approximation can be in error by more than 50\%.  Note also that
  the effective $Q$ parameter is 20--50\% larger than in the infinitesimally
  thin case.  Thus the effect of thickness is important and should be taken
  into account when analysing the stability of galactic discs.
\end{itemize}

\section*{ACKNOWLEDGMENTS}

We are very grateful to Oscar Agertz, Giuseppe Bertin, Andreas Burkert, Bruce
Elmegreen, Kambiz Fathi, Jay Gallagher, Volker Hoffmann and Chanda Jog for
useful discussions.  ABR thanks the warm hospitality of both the Department
of Physics at the University of Gothenburg and the Department of Fundamental
Physics at Chalmers.

\bsp

\label{lastpage}

\end{document}